\documentstyle[12pt]{article}        % Simple LaTeX, use with DINA4 as follows
\newlength{\dinwidth}                       
\newlength{\dinmargin}                      
\def\lsim{\mathrel{\rlap{\lower4pt\hbox{\hskip1pt$\sim$}}
    \raise1pt\hbox{$<$}}}                % less than or approx. symbol
\def\gsim{\mathrel{\rlap{\lower4pt\hbox{\hskip1pt$\sim$}}
    \raise1pt\hbox{$>$}}}                % greater than or approx. symbol
\input{epsfig.sty}

\newcommand{\be}{\begin{eqnarray}}
\newcommand{\ee}{\end{eqnarray}}

\begin{document}

\Huge{\noindent{Istituto\\Nazionale\\Fisica\\Nucleare}}

\vspace{-3.9cm}

\Large{\rightline{Sezione SANIT\`{A}}}
\normalsize{}
\rightline{Istituto Superiore di Sanit\`{a}}
\rightline{Viale Regina Elena 299}
\rightline{I-00161 Roma, Italy}

\vspace{0.65cm}

\rightline{INFN-ISS 96/6}
\rightline{August 1996}

\vspace{2cm}

\begin{center}

\Large{TAGGED NUCLEAR STRUCTURE FUNCTIONS\\ WITH $HERMES$}
\footnote{To appear in the Proceedings of the Workshop on {\em Future
Physics at HERA}, DESY (Germany), September 1995 to May 1996.}\\

\vspace{1cm}

\large{Silvano Simula}\\

\vspace{0.5cm}

\normalsize{Istituto Nazionale di Fisica Nucleare, Sezione Sanit\`{a},\\
Viale Regina Elena 299, I-00161 Roma, Italy}

\end{center}

\vspace{1cm}

\begin{abstract}

The production of slow nucleons in semi-inclusive deep inelastic electron
scattering off nuclei, $A(e, e'N)X$, is analyzed for kinematical conditions
accessible at $HERA$ with the $HERMES$ detector. The sensitivity of the
semi-inclusive cross section to possible medium-dependent modifications of the
nucleon structure function is illustrated.

\end{abstract}

\newpage

\vspace*{10cm}

\newpage

\pagestyle{plain}

\vspace*{1cm}
\begin{center}  \begin{Large} \begin{bf}
Tagged nuclear structure functions with $HERMES$\\
  \end{bf}  \end{Large}
  \vspace*{5mm}
  \begin{large}
Silvano SIMULA\\ 
  \end{large}
\end{center}
~~INFN, Sezione Sanit\`{a}, Viale Regina Elena 299, I-00161 Roma, Italy\\
\begin{quotation}
\noindent
{\bf Abstract:}
The production of slow nucleons in semi-inclusive deep inelastic electron
scattering off nuclei, $A(e, e'N)X$, is analyzed for kinematical conditions
accessible at $HERA$ with the $HERMES$ detector. The sensitivity of the
semi-inclusive cross section to possible medium-dependent modifications of the
nucleon structure function is illustrated.
\end{quotation}

\vspace{0.5cm}

\noindent The investigation of deep inelastic scattering ($DIS$) of leptons off
the nucleon and nuclei is a relevant part of the experimental activity proposed
both at present electron facilities, like $HERA$ and $TJNAF$, and at possible
future ones, like $ELFE$ and $GSI$. As is well known, existing inclusive $DIS$
data \cite{ARN94} have provided a wealth of information on quark and gluon
distributions in the nucleon and nuclei. However, important questions,
concerning, e.g., the mechanism of quark and gluon confinement as well as the
origin of the $EMC$ effect, are still awaiting for more clear-cut answers. To
this end, the investigation of semi-inclusive $DIS$ processes is expected to be
of great relevance. Recently, in \cite{CS93}-\cite{SIM96} the production of
slow nucleons\footnote{By slow nucleons we mean nucleons with momentum up to
$\sim 0.7 ~ GeV/c$ in a frame where the target is at rest (lab system).} in
semi-inclusive $DIS$ processes off nuclei, $A(\ell, \ell'N)X$, has been analyzed
within the so-called spectator mechanism, according to which, after lepton
interaction with a quark belonging to a nucleon of a correlated nucleon-nucleon
($NN$) pair, the recoiling nucleon is emitted and detected in coincidence with
the scattered lepton. The basic idea is that the momentum of the recoiling
nucleon carries information on the momentum of the struck nucleon before lepton
interaction, allowing to tag the structure function of a nucleon bound in a
nucleus. In Ref. \cite{SIM96} the semi-inclusive reaction $^2H(\ell, \ell'N)X$
has been analyzed, showing that the experimental investigation of this process
can be an effective tool to get information on the neutron structure function
as well as on the neutron to proton structure function ratio. The aim of this
contribution is to show that, at kinematical conditions accessible at $HERA$
with $HERMES$, the semi-inclusive cross section of the process $A(e, e'N)X$, for
$A > 2$, exhibits an appreciable sensitivity to possible medium-dependent
modifications of the nucleon structure function. 

\indent In case of electron scattering the semi-inclusive cross section reads as
follows
 \be
    {d^4 \sigma \over dE_{e'} ~ d\Omega_{e'} ~ dE_2 ~ d\Omega_2} = p_2 ~ E_2 ~
    \sigma_{Mott} ~ \sum_{i=L,T,LT,TT} ~ V_i(x, Q^2) ~ W^A_i(x, Q^2, \vec{p}_2)
    \label{1}
 \ee
where $x = Q^2 / 2M\nu$ is the Bjorken variable; $Q^2 = - q^2 = \vec{q}^2 -
{\nu}^2 > 0$ is the squared four-momentum transfer; $V_i$ is a kinematical
factor; $W_i^A$ is the semi-inclusive nuclear response; $\vec{p}_2$ is the
momentum of the detected nucleon and $E_2 \equiv \sqrt{M^2 + p_2^2}$ its energy
($p_2 \equiv |\vec{p}_2|$).

\indent Let us consider the process in which a virtual photon interacts with a
nucleon of a correlated $NN$ pair, and the recoiling nucleon is emitted and
detected in coincidence with the scattered electron. Within the impulse
approximation and in the Bjorken limit, the semi-inclusive nuclear structure
function $F_2^A(x, \vec{p}_2)$ is given by the following convolution formula
(cf. \cite{CS93})
 \be
    F_2^A(x, \vec{p}_2) & = & M \sum_{N_1 = n, p} Z_{N_1} \int_x^{{M_A \over 
    M} - z_2} dz_1 ~ z_1 ~ F^{N_1}_2({x \over z_1}) ~ \int d \vec{k}_{c.m.}
    ~ dE^{(2)} \nonumber \\
    & & P_{N_1N_2}(\vec{k}_{c.m.} - \vec{p}_2, \vec{p}_2, E^{(2)}) ~ \delta(M_A
    - M(z_1 + z_2) - M^f_{A-2} z_{A-2})
    \label{2}
 \ee
where $Z_{p(n)}$ is the number of protons (neutrons); $\vec{k}_1$ and
$\vec{k}_2$ are initial nucleon momenta in the lab system before interaction
with c.m. momentum $\vec{k}_{c.m.} =\vec{k}_1 + \vec{k}_2$; $\vec{p}_1 = 
\vec{k}_1 + \vec{q}$ and $\vec{p}_2 = \vec{k}_2$ are nucleon momenta in the
final state; $F_2^N$ is the structure function of the struck nucleon. In Eq.
(\ref{2}), $x / z_1$ is the Bjorken variable of the struck nucleon having
initial light-cone momentum $z_1 = k^+_1 / M$; $z_2 = (E_2 -  p_2 \cos \theta_2)
/ M$  is the experimentally measurable light-cone momentum of the detected
nucleon ($\theta_2$ is the detection angle with respect to $\vec{q}$); $z_{A-2}
= (\sqrt{(M_{A-2}^f)^2 + k_{c.m.}^2} + (k_{c.m.})_{\|}) /  M^f_{A-2}$ is the
light-cone momentum of the residual (A-2)-nucleon system with final mass $M^f
_{A-2} = M_{A-2} + E^*_{A-2}$ and intrinsic excitation energy $E^*_{A-2}$. The
relevant nuclear quantity in (\ref{2}) is the two-nucleon spectral function
$P_{N_1N_2}$, which represents the joint probability to find in a nucleus two
nucleons with momenta $\vec{k}_1$ and $\vec{k}_2$ and removal energy $E^{(2)}$.
For deuteron it simply reduces to the nucleon momentum distribution and for
$^3He$ to the square of the wave function in momentum space, times the removal
energy delta function $\delta(E^{(2)} - E^{(2)}_{thr})$, with $E^{(2)}_{thr} =
2M + M_{A-2} - M_A$ being the two-nucleon break-up threshold. In case of $^4He$
and heavier nuclei, the two-nucleon spectral function is not yet available in
the exact form; however, realistic models taking into account those features of
the two-nucleon spectral function which are relevant in the study of
semi-inclusive $DIS$ processes, have been developed \cite{CS93}-\cite{CS95}. In
this contribution the $2NC$ model of Ref. \cite{2NC}, where the c.m. motion of
the correlated pair is properly taken into  account, is adopted, viz.
 \be
    P_{N_1N_2}(\vec{k}_1, \vec{k}_2, E^{(2)}) = n^{rel}_{N_1N_2}(|\vec{k}_1 -
    \vec{k}_2| / 2) ~ n^{c.m.}_{N_1N_2}(|\vec{k}_1 + \vec{k}_2|) ~ \delta(
    E^{(2)} - E^{(2)}_{thr})
    \label{3}
 \ee
where $n^{rel}_{N_1N_2}$ and $n^{c.m.}_{N_1N_2}$ are the momentum distribution
of the relative and c.m. motion of the correlated $N_1N_2$ pair, respectively.
We point out that the $2NC$ model reproduces the high momentum and high removal
energy components of the single-nucleon spectral function of $^3He$ and nuclear
matter, calculated using many-body approaches, as well as the high momentum
part of the single-nucleon momentum distribution of light and complex nuclei
(see \cite{2NC}). Therefore, in the kinematical region $0.3 ~ GeV/c \lsim p_2
\lsim 0.7 ~ GeV/c$ for the momentum of the recoiling nucleon, the
non-relativistic description (\ref{3}) of the nuclear structure is expected to
be well grounded. Using the $2NC$ model, the nuclear effects on the energy and
angular distributions of the nucleons produced in semi-inclusive $A(e, e'N)X$
processes have been extensively investigated \cite{CS93}-\cite{CS95}, showing
that backward emission is strongly enhanced when the effects due to the c.m.
motion of the correlated pair are taken into account.

\indent Besides the spectator mechanism, there are other reaction mechanisms
which could lead to forward as well as backward nucleon emission, like, e.g.,
the so-called target fragmentation of the struck nucleon (see \cite{CS93}) and
the hadronization processes following lepton interactions with possible
six-quark ($6q$) cluster configurations at short $NN$ separations (see
\cite{SIM94,CS95}). Furthermore, it should be considered that within the
spectator mechanism the virtual boson can be elastically absorbed by the struck
nucleon. The contribution from this process, which involves the nucleon form
factor instead of the nucleon structure function, vanishes in the Bjorken
limit, but it can affect the semi-inclusive cross section at finite values of
$Q^2$ and for $x \to 1$. In what follows, we will refer to such a process as the
quasi-elastic ($QE$) contamination. 

\indent The semi-inclusive cross section of the process $^{12}C(e, e'p)X$ has
been calculated including in (\ref{1}) all the nuclear response functions and
considering electron kinematical conditions accessible at $HERA$ with
$HERMES$ (i.e., $E_e = 30 ~ GeV$ and $Q^2$ in the range $5 \div 15 ~
(GeV/c)^2$). The relative and c.m. momentum distributions adopted in the
calculations are taken from \cite{2NC}. As for the nucleon structure function,
the parametrization of the $SLAC$ data of Ref. \cite{SLAC} has been considered.
In case of backward proton emission the results obtained for the contributions
resulting from the spectator mechanism, the (above-mentioned) hadronization
processes and the $QE$ contamination are separately shown in Fig. 1 as a
function of the kinetic energy $T_2$ of the detected nucleon. It can clearly be
seen that for $50 ~ MeV \lsim T_2 \lsim 250 ~ MeV$ (corresponding to $0.3 ~
GeV/c \lsim p_2 \lsim 0.7 ~ GeV/c$) backward nucleon emission is mainly
governed by the spectator mechanism (cf. \cite{CS93}-\cite{CS95}). Therefore,
in what follows, we will limit ourselves to the case of backward nucleon
emission.

\begin{figure}[t]

\epsfig{file=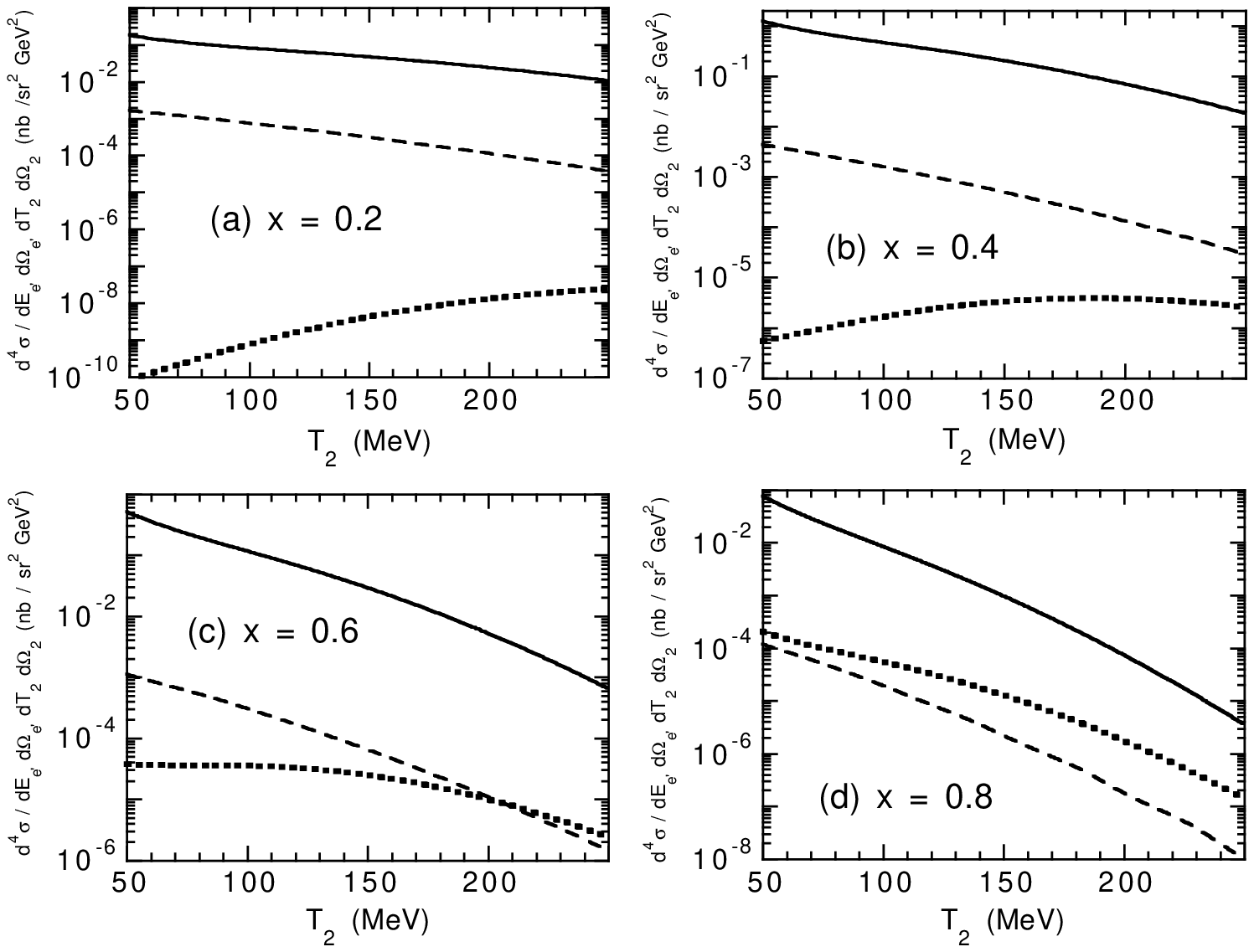}

\vspace{-17.0cm}

\parbox{0.25cm} \ $~$\ \parbox{16cm}{\small {\noindent \it Fig. 1. The
semi-inclusive cross section (\ref{1}) of the process $^{12}C(e, e'p)X$,
evaluated at $Q^2 = 10 ~ (GeV/c)^2$ and for backward proton emission at
$\theta_2 = 140^o$, versus the kinetic energy $T_2$ of the detected nucleon at
various values of the Bjorken variable $x$. The solid lines are the results
obtained within the spectator mechanism. The dashed lines correspond to the
proton emission arising both from the target fragmentation of the struck
nucleon, evaluated as in \cite{CS93}, and from virtual photon absorption on
$6q$ cluster configurations, evaluated as in \cite{SIM94,CS95}. The dotted lines
are the contribution from the $QE$ contamination (see text).}}

\end{figure}

\indent In \cite{CS93} a ratio of semi-inclusive cross sections, evaluated at
different values of $x$ but keeping fixed both $Q^2$ and the nucleon
kinematical variables, has been introduced, viz.
 \be
    R_1(x_0, x; Q^2, \vec{p}_2) \equiv d^4 \sigma (x, Q^2, \vec{p}_2) ~ / ~
    d^4 \sigma (x_0, Q^2, \vec{p}_2)
    \label{4}
 \ee
The ratio $R_1$ turns out to be almost independent of the effects due to the
rescattering of the recoiling nucleon with the residual ($A-2$)-nucleon system
(cf. \cite{CS93}); such an important feature is mainly due to the fact that the
nucleon kinematical variables are the same in the numerator and denominator of
the cross section ratio $R_1$. In order to investigate the sensitivity of $R_1$
to possible medium-dependent modifications of the nucleon structure function,
three models available in the literature have been considered. In the first one
\cite{FS81} the valence-quark distributions in the nucleon are expected to be
suppressed when the nucleon is bound in a nucleus, since point-like
configurations ($plc$) in the nucleon should interact weaker in the medium with
respect to normal-size configurations. In \cite{FS81} the suppression factor is
expected to be a function of the momentum of the struck nucleon. The second and
third models are $Q^2$-rescaling models \cite{CJRR85,DT86}, where the rescaling
is driven by nucleon swelling \cite{CJRR85} or by the off-shellness of the
struck nucleon \cite{DT86}. The results obtained are reported in Fig. 2 in terms
of the ratio of the quantity $R_1$, evaluated using the medium-modified nucleon
structure function, to the quantity $R_1$, calculated with the free $F_2^N$. It
can clearly be seen that the ratio $R_1$ is remarkably sensitive to possible
deformations of the nucleon structure function.

\begin{figure}[t]

\parbox{7cm}{\epsfig{file=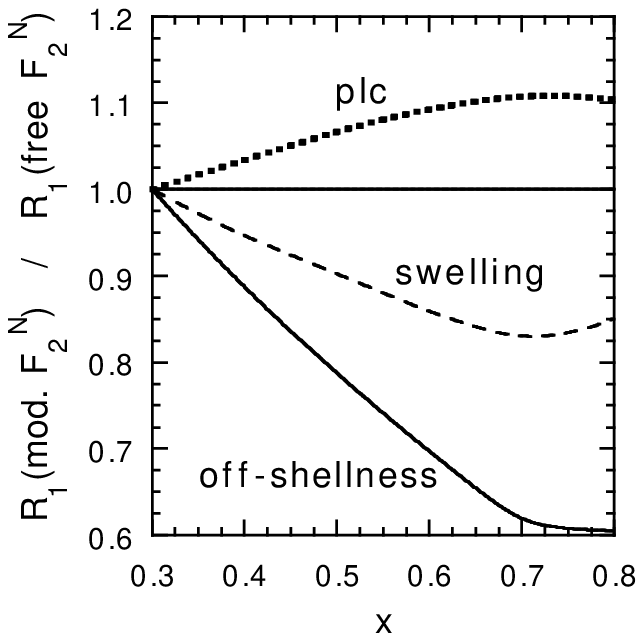} \vspace{-23.0cm}}
\ $~~~~$ \
\parbox{8cm}{\small {\noindent \it Fig. 2. The ratio of the quantity $R_1(x_0 =
0.3, x; Q^2, \vec{p}_2)$ (Eq. (\ref{4})), calculated using medium-modified and
free nucleon structure functions. The calculations have been performed in case
of the process $^{12}C(e, e'p)X$ at $Q^2 = 10 ~ (GeV/c)^2$, $p_2 = 0.4 ~ GeV/c$
and for backward proton kinematics ($\theta_2 = 140^o$). The dotted, dashed and
solid lines correspond to the models of Refs. \cite{FS81}, \cite{CJRR85} and
\cite{DT86}, respectively.}}

\end{figure}

\indent Another useful cross section ratio can be defined as
 \be
    R_2(Q_0^2, Q^2; x, \vec{p}_2) \equiv d^4 \sigma (x, Q^2, \vec{p}_2) ~ / ~
    d^4 \sigma (x, Q_0^2, \vec{p}_2)
    \label{5}
 \ee
where both $x$ and the nucleon kinematical variables are kept fixed. The ratio
$R_2$ is expected to be mainly dominated by the $Q^2$ behaviour of the nucleon
structure function. As a matter of fact, explicit calculations show that
$R_2$ is almost independent both of $p_2$ and of the mass number $A$, when the
free nucleon structure function is adopted. Besides the three models employed
in the calculations shown in Fig. 2, a further model \cite{HT90} has been
considered. It generates both $x$- and $Q^2$-rescaling of $F_2^N$, driven by
binding effects on the energy transferred to the struck nucleon. The results are
shown in Fig. 3. It can be seen that also the ratio $R_2$ is appreciably
affected by possible deformations of the nucleon structure function and,
moreover, the $p_2$-dependence of the deviations with respect to the
predictions obtained using free $F_2^N$, could provide relevant information on
the type of medium effects on $F_2^N$.  

\begin{figure}[t]

\epsfig{file=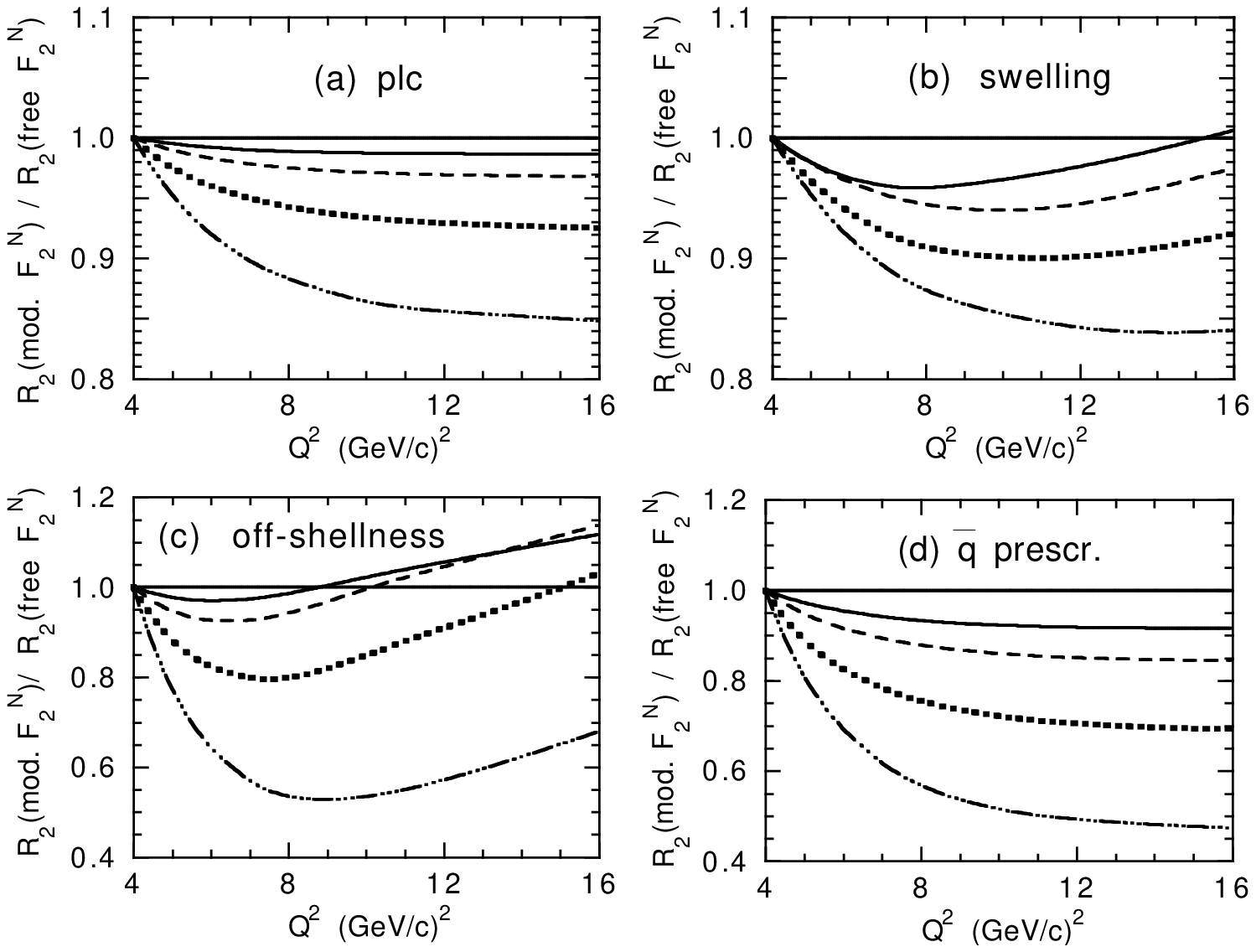}

\vspace{-17.0cm}

\parbox{0.25cm} \ $~$\ \parbox{16cm}{\small {\noindent \it Fig. 3. The ratio of
the quantity $R_2(Q_0^2 = 4 ~(GeV/c)^2, Q^2; x, \vec{p}_2)$ (Eq. (\ref{5})),
calculated using medium-modified and free nucleon structure functions. The
calculations have been performed in case of the process $^{12}C(e, e'p)X$ at $x
= 0.6$ and for backward nucleon kinematics ($\theta_2 = 140^o$). The solid,
dashed, dotted and dot-dashed lines correspond to $p_2 = 0.3, 0.4, 0.5$ and
$0.6 ~ GeV/c$, respectively. The models adopted for the description of the
medium-dependent modifications of $F_2^N$ are from \cite{FS81} (a),
\cite{CJRR85} (b), \cite{DT86} (c) and \cite{HT90} (d), respectively.}}

\end{figure}

\indent Before closing, it should be reminded that our calculations have been
performed within the assumption that the debris produced by the fragmentation
of the struck nucleon does not interact with the recoiling spectator nucleon.
Estimates of the final state interactions of the fragments in semi-inclusive
processes off the deuteron have been obtained in \cite{TN92}, suggesting that
rescattering effects should play a minor role thanks to the finite formation
time of the dressed hadrons. Moreover, backward nucleon emission is not
expected to be sensitively affected by forward-produced hadrons (cf.
\cite{BDT94}), and the effects due to the final state interactions of the
fragments are expected to cancel out (at least partially) in the cross section
ratios $R_1$ and $R_2$.

\indent In conclusion, the production of slow nucleons in semi-inclusive deep
inelastic electron scattering off nuclei, $A(e, e'N)X$, has been investigated in
kinematical regions accessible at $HERA$ with the $HERMES$ detector. It has
been shown that backward nucleon production is mainly governed by the spectator
mechanism, provided the Bjorken variable $x$ and the kinetic energy of the
detected nucleon are in the range $0.2 \div 0.8$ and $50 \div 250 ~ MeV$,
respectively. The ratios (\ref{4}) and (\ref{5}) of the semi-inclusive cross
sections, evaluated at different electron kinematics keeping fixed the nucleon
ones, exhibit an appreciable sensitivity to possible medium-dependent
modifications of the nucleon structure function.

\end{document}